\documentclass[12pt]{article}
\usepackage[dvips]{graphicx}

\usepackage{epsfig,amsmath,amssymb,verbatim,mathrsfs}

\usepackage{slashed}
\usepackage{graphicx}
\usepackage{amsmath}
\usepackage{amssymb}
\usepackage{epstopdf}
\usepackage{verbatim}
\usepackage{url}
\usepackage{amsfonts}
\usepackage{layout}
\usepackage{slashed}
\usepackage[usenames,dvipsnames]{color}
\usepackage{cancel}

\usepackage{hyperref}	

\newcommand{\eref}[1]{Eq.~(\ref{#1})}

\newcommand{\drawsquare}[2]{\hbox{%
\rule{#2pt}{#1pt}\hskip-#2pt
\rule{#1pt}{#2pt}\hskip-#1pt
\rule[#1pt]{#1pt}{#2pt}}\rule[#1pt]{#2pt}{#2pt}\hskip-#2pt
\rule{#2pt}{#1pt}}

\newcommand{\fund}{~\raisebox{-.5pt}{\drawsquare{6.5}{0.4}}~}
\newcommand{\antifund}{~\overline{\raisebox{-.5pt}{\drawsquare{6.5}{0.4}}}~}

\newcommand{\beq}{\begin{equation}}
\newcommand{\eeq}{\end{equation}}
\newcommand{\bea}{\begin{eqnarray}}
\newcommand{\eea}{\end{eqnarray}}

\newcommand{\vev}[1]{\langle #1 \rangle}

\newcommand{\mr}[1]{\mathrm{#1}}

\newcommand{\be}{\begin{equation}}
\newcommand{\ee}{\end{equation}}

\newcommand{\ben}{\begin{equation*}}
\newcommand{\een}{\end{equation*}}

\newcommand{\mc}{\mathcal}
\renewcommand{\t}{\tilde}

\begin{document}

\setlength{\baselineskip}{0.2in}


\begin{titlepage}
\noindent
\begin{flushright}
SLAC-PUB-14924
\end{flushright}
\vspace{1cm}

\begin{center}
  \begin{Large}
    \begin{bf}

An Attractor for Natural Supersymmetry
     \end{bf}
  \end{Large}
\end{center}
%

\begin{center}

{\bf Timothy Cohen, Anson Hook, and Gonzalo Torroba}\\
\vspace{.5cm}
  \begin{it}
Theory Group, SLAC National Accelerator Laboratory, \\
2575 Sand Hill Rd, Menlo Park, CA 94025\\
\vspace{0.5cm}
\end{it}
%
%

\end{center}


\begin{abstract}
We propose an attractor mechanism which generates the ``more minimal" supersymmetric standard model from a broad class of supersymmetry breaking boundary conditions.  The hierarchies in the fermion masses and mixings are produced by the same dynamics and a natural weak scale results from gaugino mediation. These features arise from augmenting the standard model with a new $SU(3)$ gauge group under which only the third generation quarks are charged. The theory flows to a strongly interacting fixed point which induces a negative anomalous dimension for the third generation quarks and a positive anomalous dimension for the Higgs.  As a result, a split-family natural spectrum and the flavor hierarchies are dynamically generated. 
\end{abstract}

\vspace{1cm}

\end{titlepage}

\noindent

\def\thefootnote{\arabic{footnote}}
\setcounter{footnote}{0}

\setcounter{page}{2} 

\newpage
\section{Introduction and Summary\label{sec:introduction}}

The stability of the electroweak scale and the hierarchical structure of the fermion masses and mixing angles are two of the central mysteries of the Standard Model (SM).  It is possible that these puzzles are explained by the same underlying mechanism.   One approach is to supersymmetrize the SM and augment it with a new strongly interacting gauge theory.  Supersymmetry tames the quadratically divergent contributions to the Higgs mass while the strong dynamics can yield a parametric suppression of the first and second generation Yukawa couplings.  Various realizations of this possibility have been proposed so far, including single sector models~\cite{ArkaniHamed:1997fq}, models of superconformal flavor~\cite{Nelson:2000sn,Poland:2009yb}, warped extra-dimensional realizations~\cite{Gherghetta:2000qt}, and theories based on deconstruction~\cite{Craig:2011yk}.\footnote{For some other models which connect the supersymmetry breaking spectrum and flavor, see \cite{OtherFlavorAndSUSYBreakingModels}.}

For some of these constructions, the dynamics which gives
rise to the flavor textures also produces an inverted squark hierarchy, where the lightest SM fermions
have the heaviest sfermion partners. This provides a microscopic realization of the ``more
minimal'' supersymmetric SM of~\cite{minimal}, which was motivated by considerations of naturalness and flavor constraints. The phenomenology of these models has been studied thoroughly in \emph{e.g.}~\cite{SplitFamilyPheno}. Furthermore, the recent LHC bounds on first and second generation squark masses~\cite{Aad:2011ib} together with attempts to minimize fine tuning have reinvigorated interest in the phenomenology and collider signatures of such ``natural supersymmetry'' spectra~\cite{Essig:2011qg,Papucci:2011wy}.  

In this work, we will present a new model to explain the flavor hierarchies
which simultaneously yields the natural supersymmetry spectrum and radiative electroweak symmetry breaking (REWSB). This will be accomplished by adding a new strongly coupled conformal sector to the minimal supersymmetric standard model (MSSM). We will show that starting from rather generic supersymmetry breaking boundary conditions (with some assumptions on certain approximate global symmetries), the infrared theory after escaping from the conformal regime is the more minimal supersymmetric SM.

The MSSM fields are weakly coupled both in the UV and in the IR.  The conformal dynamics will generate order one negative anomalous dimensions for the third generation fields once the theory becomes strongly coupled.\footnote{This differs from previously studied constructions, which relied on large positive anomalous dimensions for the first two SM generations.  This can result from compositeness or localization in the IR region of a Randall-Sundrum throat.}  Negative anomalous dimensions are only possible if the third generation is charged under this new gauge group --- the unitarity bound on dimensions only applies to gauge invariant operators.  The third generation Yukawa couplings are marginal operators in the conformal field theory (CFT).  These marginal Yukawa couplings will induce a large positive anomalous dimension for the Higgs field.  Hence, the remaining Yukawas become irrelevant deformations.   It will be shown that this structure can lead to viable flavor hierarchies.  Additionally, the strong dynamics will suppress soft masses for the third generation squarks and Higgs fields.  Below the exit scale, these will be regenerated by gaugino mediation \cite{Kaplan:1999ac}.  The model acts as an attractor for the more minimal supersymmetry spectrum and REWSB.

\begin{figure}[htb]
\begin{center}     
\includegraphics[height=4.cm]{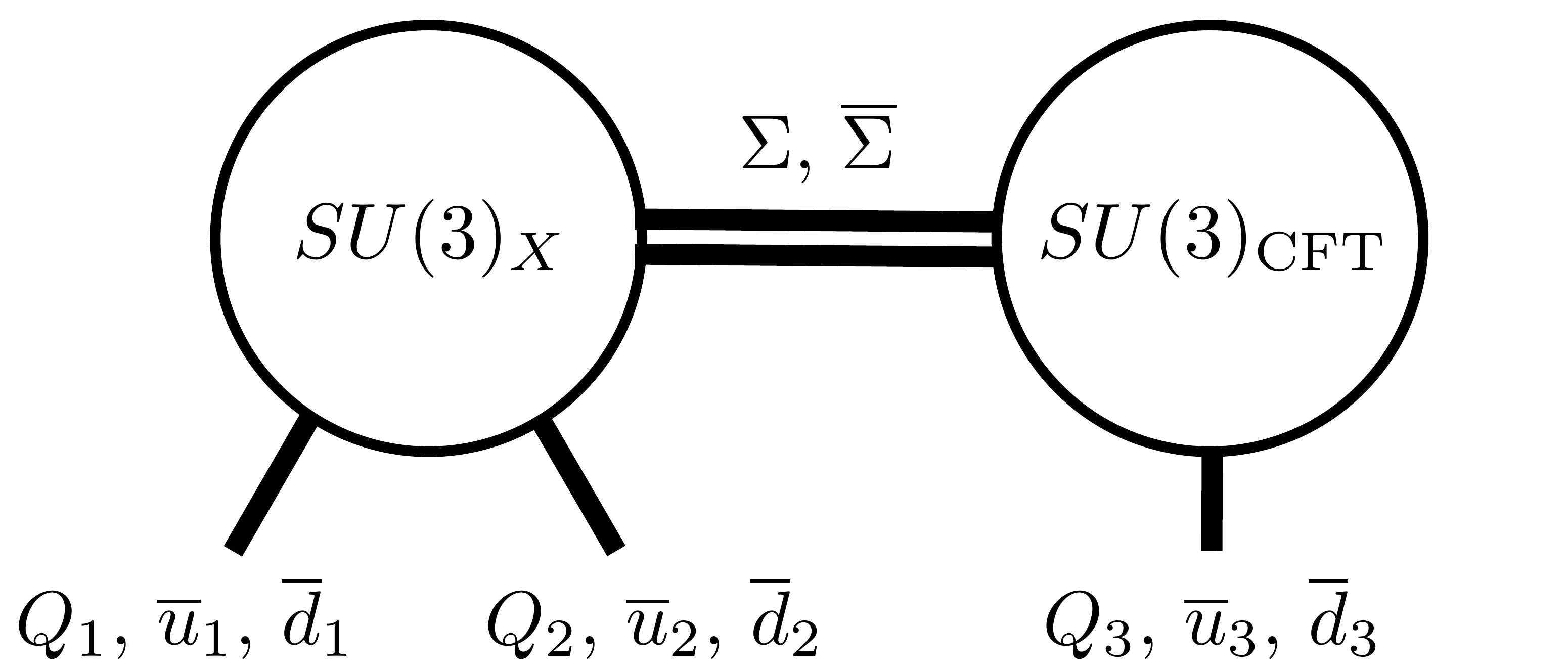}
\caption{\small{The model presented here is given by an $SU(3)_X \times SU(3)_\mr{CFT}$ quiver gauge theory. The node $SU(3)_\mr{CFT}$ flows to an interacting fixed point and provides the necessary dynamics for generating flavor and an attractor mechanism for natural supersymmetry. $SU(3)_X$ is IR free. The bifundamental link fields $\Sigma$ and $\overline{\Sigma}$ break the group to the diagonal visible $SU(3)_C$, providing an exit from the conformal regime.}}
\label{fig:quiver}
\end{center}
\end{figure}

The goal of this work is to analyze the simplest realization of this mechanism and its main dynamical consequences. The gauge group is $SU(3)_\text{CFT} \times SU(3)_X \times SU(2)_W \times U(1)_Y$ where $SU(3)_\mr{CFT}$ will flow to a strongly coupled fixed point, $SU(3)_X$ is weakly coupled, and $SU(2)_W\times U(1)_Y$ are as in the MSSM. The third generation quark superfields transform under $SU(3)_\text{CFT}$.  The first and second generations transform under $SU(3)_X$. The $SU(3)$ groups are connected by bifundamental `link' fields. With this matter content, $SU(3)_\text{CFT}$ is in the conformal window~\cite{Seiberg:1994pq}.  The link fields eventually acquire a nonzero expectation value causing an exit from the conformal regime; this also breaks $SU(3)_\text{CFT} \times SU(3)_X \to SU(3)_C$, giving rise the visible color interactions.\footnote{The super top color model of \cite{Fukushima:2010pm} utilizes a similar group structure and matter content.  However, unlike models of top color (see \cite{Hill:2002ap} for a review), the mechanism studied in this work does not utilize top condensation to break electroweak symmetry.}   This structure is summarized in Fig.~\ref{fig:quiver}.

\vskip 1cm

The rest of this paper is organized as follows.  \S \ref{sec:TheModel} describes the basic mechanism and its implications for the spectrum of soft masses and the flavor hierarchies. In \S \ref{sec:SpectrumAndPheno} we discuss the low energy phenomenology, including some general remarks about the spectrum and the Higgs sector.  We also provide some concrete example spectra. Our conclusions and future directions are presented in \S \ref{sec:concl}.

\section{The Model}\label{sec:TheModel}

We begin by describing the model of Fig.~\ref{fig:quiver}. For simplicity we will ignore the leptons, which do not affect our discussion other than to ensure anomaly cancellation. The matter content and charge assignments are given in Table \ref{tab:ParticleContent}. The third generation quarks are charged under $SU(3)_\text{CFT}$, while the first two generations are charged under $SU(3)_X$. The bi-fundamental link fields are denoted by $\Sigma$ and $\overline \Sigma$.  The field $A$ is an adjoint plus a singlet of $SU(3)_X$.  The superpotential will be chosen so that the $F$-term for $A$ forces $\vev{\Sigma}= \vev{\overline{\Sigma}} \neq 0$.  This will cause an escape from the conformal regime while also giving masses to the bi-fundamentals.  The superpotential contains the following relevant terms:
\be\label{eq:Superpotential}
W\supset  Q_3\, H_u\, \overline{u}_3 + Q_3\, H_d \, \overline{d}_3 + \Sigma\,A\, \overline{\Sigma} + W_{\leavevmode\cancel{U(1)}}\,.
\ee
Contractions over gauge indices are implicit.  $W_{\leavevmode\cancel{U(1)}}$ will be instrumental in breaking some of the Abelian symmetries which can spoil the desired low energy spectrum. We will discuss this term in detail below.

\begin{table}
\begin{center}
\begin{tabular}{c|c|ccc}
&$SU(3)_\mr{CFT}$&$SU(3)_X$&$SU(2)_{W}$&$U(1)_Y$  \\
\hline
\hline
&&&\\[-12pt]
$Q_3$&$\fund$&$1$&$\fund$&$1/6$  \\
$\overline{d}_3$&$\antifund$&$1$&$ 1$ &$1/3$ \\
$\overline{u}_3$&$\antifund$&$1$&$1$&$\:\!\!\!\!\!\!-2/3$  \\
$H_u$&$1$&$1$&$\fund$&$1/2$\\
$H_d$&$1$&$1$&$\fund$&$\:\!\!\!\!\!\!-1/2$ \\
\hline
$\Sigma$&$\fund$&$\antifund$&$1$&$0$ \\
$\overline{\Sigma}$&$\antifund$&$\fund$&$1$&$0$ \\
$A$&$1$&$1 + \mr{adj}$&$1$&$0$ \\
\hline
$Q_{2,1}$&$1$&$\fund$&$\fund$&$1/6$  \\
$\overline{d}_{2,1}$&$1$&$\antifund$&$ 1$ &$1/3$ \\
$\overline{u}_{2,1}$&$1$&$\antifund$&$1$&$\:\!\!\!\!\!\!-2/3$
\end{tabular}
\end{center}
\caption{\small The particle content and charge assignments for the MSSM quark and CFT sectors.  The subscripts denote generation assignments, and the leptons are charged as in the MSSM.  The visible color gauge group is a diagonal subgroup of $SU(3)_\mr{CFT} \times SU(3)_X$.}
\label{tab:ParticleContent}
\end{table}

With this matter content, $SU(3)_\text{CFT}$ has five flavors and flows to a strongly interacting superconformal fixed point in the IR. The crossover scale below which this theory becomes strong is denoted by $\Lambda_\text{CFT}$. The remaining gauge groups are IR free and act as spectators to this strong dynamics. A crucial property of the model is that the third generation Yukawa couplings appear as relevant interactions in the CFT. The Higgs fields will then also be part of the CFT --- they will receive a positive anomalous dimension. These couplings, as well as the rest of the interactions in \eref{eq:Superpotential}, will naturally flow to order one values below $\Lambda_\text{CFT}$. In contrast, the remaining Yukawas will arise as irrelevant deformations, resulting in a flavor hierarchy between the third and first two generations.

If we do not add extra fields, this matter content spoils gauge coupling unification. However, there are no issues with Landau poles up to the GUT scale, and one could imagine UV completing the model using full $SU(5)$ representations. We will come back to this point briefly in \S \ref{sec:concl}, while here we continue to focus on this minimal realization.

\begin{figure}[htb]
\begin{center}     
\includegraphics[height=3.7cm]{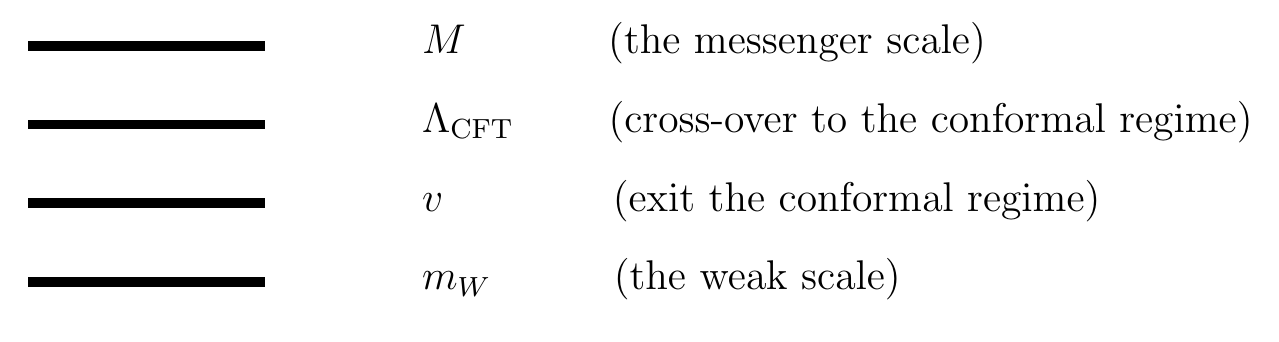}
 \caption{\small The relevant scales for our model.} \label{fig:Scales}
\end{center}
\end{figure}

The energy scales in our model are as follows, see Fig.~\ref{fig:Scales}.  At the messenger scale $M$, soft supersymmetry breaking operators are generated. The supersymmetry breaking mechanism and mediation can be arbitrary, up to certain assumptions on global symmetries which we describe below. The scale $M$ could be above or below $\Lambda_\text{CFT}$, but the physical soft masses should be smaller than $\Lambda_\text{CFT}$ so that the superconformal dynamics dominate.  At a scale $v < \Lambda_\text{CFT}$, we exit the CFT regime. This is done supersymmetrically by adding
\be\label{eq:Wexit}
W \supset - v^2\, \mr{Tr}A\,
\ee
to \eref{eq:Superpotential}. This new scale can be generated dynamically as explained in~\cite{Dine:2006gm}. The link fields acquire an expectation value $\langle \Sigma \overline{\Sigma} \rangle =v^2$, which breaks $SU(3)_\mr{CFT}\times SU(3)_X \rightarrow SU(3)_C$. The visible gauge coupling becomes
\be
\frac{1}{g_C^2}=\frac{1}{g_X^2}+\frac{1}{g_\text{CFT}^2}\,,
\ee
which is dominated by $g_X^2\simeq g_C^2$. We assume that the exit from the conformal regime happens quickly, such that at energy scales $E < v$ a perturbative description is valid.  As we show below, the weak scale $m_W < v$ is radiatively generated. We note that, in contrast with composite models, here the MSSM fields are weakly coupled both in the UV (above $\Lambda_\text{CFT}$) and in the IR (below the exit scale).

\subsection{An Attractor for Natural Supersymmetry}\label{sec:AttractorForNaturalSUSY}

In this section, we will analyze the conformal regime and how it affects the soft masses. For more details, see~\cite{Nelson:2001mq} and the references therein.  We will first neglect the effects from the weakly interacting gauge groups and the first two generations. This corresponds to setting $g_\mr{SM} \to 0$ and ignoring mixings from Yukawa couplings. We will then show that such effects 
amount to small finite corrections.

Our main dynamical assumption is that the fixed point is stable, which means that small perturbations of the couplings away from their fixed point value are irrelevant.  Equivalently, the matrix $\partial \beta_i/\partial y_j$ must be positive definite, where $y_i$ are the couplings of the theory and $\beta_i$ are the corresponding beta functions.  With this assumption, all the physical couplings flow to their fixed point values, and their higher $\theta$ components flow to zero.  This can be seen by promoting the couplings to background superfields. One implication is that all soft supersymmetry breaking terms associated with relevant couplings are highly suppressed by the conformal dynamics.\footnote{For a model which uses this mechanism to suppress the Higgs soft mass, see \cite{Kobayashi:2004pu}.} 

To understand the consequences for our model, consider a relevant superpotential interaction
\bea
W \supset \lambda\, \prod_i \,\Phi_i^{n_i},
\eea
for superfields $\Phi_i$ and positive integers $n_i$.  The physical coupling is
\beq
\lambda_\mr{phys} = \lambda\,  \prod_i \left(Z_{\Phi_i}^{n_i}\right)^{-1/2}
\eeq
where the $Z_{\Phi_i}$ are the wave function renormalizations for the superfields $\Phi_i$ and encode the soft masses as their $\theta^4$ components.  As $\lambda_\mr{phys}$ flows to its fixed point value, its $\theta^4$ component flows to zero.  Equivalently, this implies that the combination of soft masses
\bea\label{eq:soft1}
\sum_i n_i \,\widetilde{m}_i^2
\eea
flows to zero at the fixed point, where $\widetilde{m}_i$ is the soft mass for $\Phi_i$. Since the $\theta^2$ component also flows to zero, the same conclusion holds for the $a$-terms. 

Similarly, promoting the gauge coupling to a superfield implies that the CFT gaugino mass and
\bea
\label{eq:soft2}
\sum_r \text{dim}(r)\, T_r\, \widetilde{m}_r^2
\eea
are also suppressed by the CFT dynamics. Here the field $\Phi_r$ has index $T_r$ under the gauge group, \emph{e.g.}~$T(\Box)=1/2$, and $\text{dim}(r)$ is the dimension of $\Phi_r$ for a fixed gauge index. 

As we mentioned above, we assume that the CFT is IR attractive, which means that the eigenvalues $\lambda_i$ of the matrix $\partial \beta_i/ g_j$ are positive and, generically at a strongly coupled fixed point, order one. The previous soft parameters are then suppressed by a power-law $(\frac{\mu}{\Lambda_\text{CFT}})^{\lambda_i}$, where $\mu$ is the RG scale. This effect can be seen explicitly in weakly coupled examples such as the Wilson-Fisher fixed point. Below we will take into account the small contributions from the perturbative SM couplings. 

On the other hand, due to the non-renormalization of conserved currents, combinations of soft masses proportional to conserved $U(1)$ symmetries,
\be\label{eq:massesProtectedByU1}
\sum_i  \text{dim}(i) \,q_i\, \widetilde{m}_i^2 
\ee
are not renormalized by the strong dynamics. Here $q_i$ denotes the $U(1)$ charge. This effect will be an important constraint on the viability of obtaining the more minimal supersymmetric SM.

Our goal is to use the conformal dynamics to suppress the soft masses for the third
generation squarks and Higgs fields, which is essentially the idea of conformal sequestering~\cite{Schmaltz:2006qs}.  We must explicitly break some of the non-anomalous global $U(1)$ symmetries.  Otherwise, \eref{eq:massesProtectedByU1} shows that they would lead to unsuppressed tachyonic soft masses. We accomplish this with
the term $W_{\leavevmode\cancel{U(1)}}$ in the superpotential of \eref{eq:Superpotential}. As a concrete example, let us investigate a specific choice:
\be\label{eq:ExplicitWU1}
W_{\leavevmode\cancel{U(1)}}= (Q_3 \,\overline u_3)(Q_3\, \overline d_3)\,.
\ee
In this theory, the superconformal $R$-charges are uniquely determined in terms of symmetries and anomaly cancellation. The anomaly-free Abelian symmetries are given in Table \ref{tab:U1Charges}.

\begin{table}[h!!!]
\begin{center}
\begin{tabular}{c|ccc|c}
&$U(1)_1$&$U(1)_2$&$U(1)_3$&$U(1)_R$ \\
\hline
\hline
&&&&\\[-12pt]
$Q_3$ & $1$ & $0$ & $0$ & $1/2$\\
$\overline u_3$ & $\:\!\!\!\!\!\!-1$ & $\!\!\!\!\!-1$ & $0$ & $1/2$\\
$\overline d_3$ & $\:\!\!\!\!\!\!-1$ & $1$ & $0$ & $1/2$\\
$H_u$ & $0$ & $1$ & $0$ & $1$\\
$H_d$ & $0$ & $\:\!\!\!\!\!\!-1$ & $0$ & $1$\\
\hline
$\Sigma$ & $0$ & $0$ & $1$ & $1/3$\\
$\overline \Sigma$ & $0$ & $0$ & $\:\!\!\!\!\!\!-1$ & $1/3$\\
$A$ & $0$ & $0$ & $0$ & $4/3$\\
\end{tabular}
\end{center}
\caption{The global anomaly free $U(1)$ symmetries for the model given by \eref{eq:Superpotential} with the $U(1)$ breaking superpotential in \eref{eq:ExplicitWU1}.  The charge assignments for the gauged symmetries are given in Table \ref{tab:ParticleContent}.}
\label{tab:U1Charges}
\end{table}

One can verify self-consistently that all terms in the superpotential in \eref{eq:Superpotential} have $R$-charge 2 at the fixed point. The remaining first and second generation fields decouple from the strong dynamics; they are neutral under the non-$R$ symmetries and are (approximately) free fields.

It is useful to explain in more detail how the flow to this fixed point proceeds, starting from the UV free theory. In the UV, the cubic terms in the superpotential are classically marginal, but the quartic symmetry breaking term is classically irrelevant. In terms of the canonical UV fields, $W_{\leavevmode\cancel{U(1)}}= \frac{1}{M_*}(Q_3 \,\overline u_3)(Q_3\, \overline d_3)$, where $M_*$ is some large mass scale. First, consider the limit $M_* \to \infty$. The resulting theory is Supersymmetric QCD with extra singlets and cubic superpotential deformations (SSQCD). Below the strong coupling scale for this model $\Lambda_\text{SSQCD}$, the theory flows to a superconformal fixed point. This CFT can be studied using $a$-maximization~\cite{Intriligator:2003jj}; we find that the superconformal $R$-charges of $Q_3$ and $(\bar u_3, \bar d_3)$ are $2/3 \times \sqrt{2/7}$. The cubic interactions are then relevant perturbations of the free fixed point, driving the theory to the nontrivial SSQCD fixed point.

Next, we can add the quartic superpotential, taking $M_*/\Lambda_\text{SSQCD}$ large but finite. The theory first flows close to the SSQCD fixed point which we just described. According to the previous $R$-charges, in this regime the quartic operator $(Q_3 \,\overline u_3)(Q_3\, \overline d_3)$ is relevant. So for any nonzero value of $M_*/\Lambda_\text{SSQCD}$ it will drive the theory away from the SSQCD fixed point. The crossover scale $\Lambda_\text{CFT}$ at which such effects become important is of order $\Lambda_\text{CFT}^{3-4\Delta} \sim \Lambda_\text{SSQCD}^{4-4\Delta}/M_*$. Below this scale the fixed point value of the quartic coupling is order one irrespective of the initial $M_*$, and we recover the CFT with the $R$-charges given in Table \ref{tab:U1Charges}.  We note that for larger $M_*$ it takes longer to flow to this fixed point; however, since the fixed point values are always order one, the $U(1)$ symmetry breaking term will suppress the unwanted soft masses as long as $\widetilde{m}< \Lambda_\text{CFT}$.\footnote{This analysis does not conflict with the results of \cite{Green:2010da} since $(Q_3 \,\overline u_3)(Q_3\, \overline d_3)$ is not a chiral primary operator at the fixed point with superconformal $R$-charges given in Table \ref{tab:U1Charges}. We thank Dan Green for discussions on this point.}

Having analyzed the renormalization group (RG) evolution towards our fixed point, let us return to the behavior of the soft parameters.
The only combinations of soft masses which are not suppressed as the fixed point is approached are
\bea\label{eq:unsuppressed}
&&\widetilde{m}_\Sigma^2 - \widetilde{m}_{\overline{\Sigma}}^2 \nonumber\\
&&2 \,\widetilde{m}_{Q_3}^2 - \widetilde{m}_{\overline{u}_3}^2 -\widetilde{m}_{\overline{d}_3}^2 \nonumber\\
&& \widetilde{m}_{H_u}^2 -\widetilde{m}_{H_d}^2 + \widetilde{m}_{\overline{d}_3}^2- \t m_{\overline{u}_3}^2\,.
\eea
This implies that for arbitrary UV boundary conditions, the model does not fully sequester soft masses. However, if the supersymmetry breaking mechanism preserves approximate charge conjugation and custodial symmetries, then the contributions from \eref{eq:unsuppressed} are negligible at the messenger scale and are not generated by the strong dynamics. This is the case in minimal gauge mediation \cite{Giudice:1998bp}, where at the messenger scale the first difference in \eref{eq:unsuppressed} vanishes identically, while the linear combinations in the second and third lines are much smaller than each of their respective terms. These combinations can also be suppressed by going beyond minimal gauge mediation or in gravity mediation by imposing discrete symmetries.

As we noted before, this analysis neglects effects from the weakly interacting sector of the theory. The first two generations and the SM gauginos continuously feed supersymmetry breaking contributions to the CFT fields, giving rise to ``driving terms'' in the beta functions for the CFT superfield couplings. However, these supersymmetry breaking effects are much smaller than the soft masses of the first two generation sfermions and gauginos, since the CFT couples to such fields only through irrelevant interactions. Specifically, they are suppressed by loop factors and by SM gauge couplings or Yukawa interactions. These corrections will be taken into account in \S \ref{sec:SpectrumAndPheno}.

Hence, under the assumption that the supersymmetry breaking mechanism (approximately) respects the above symmetries, the strong conformal dynamics fully suppresses the soft masses of the third generation quarks and Higgs fields, up to small corrections from the weakly coupled sector. It would be interesting to modify the model to accomplish a complete sequestering without having to assume these symmetries, \emph{e.g.}~by adding new flavors and turning on different deformations. Some of these possibilities will be discussed briefly in \S \ref{sec:concl}.

Finally, at the scale $v$ we exit the conformal regime.  This happens in an approximately supersymmetric way and does not lead to appreciable finite corrections for the soft parameters. Therefore, the theory at energies below $v$ is the weakly coupled MSSM with the soft masses for the third generation fields and Higgses suppressed with respect to the first two generations and the gauginos. The third generation squark masses are then regenerated by gaugino mediation~\cite{Kaplan:1999ac}, which in turn can drive the up-type Higgs soft mass squared negative. Starting from generic supersymmetry breaking mechanisms, the CFT acts an an attractor for realizing the natural supersymmetry spectrum of the more minimal supersymmetric SM. The resulting phenomenology will be studied in \S \ref{sec:SpectrumAndPheno}.

\subsection{Generating the Flavor Hierarchies}

Starting from the pioneering work of Nelson and Strassler~\cite{Nelson:2000sn}, it has been understood how CFT dynamics can generate the flavor hierarchies at low energies from arbitrary order one Yukawas in the UV. We will explain how this works in the context of our construction.  The model presented here differs from previous models of compositeness/superconformal flavor since the third generation superfields have negative anomalous dimensions.

Above the dynamical scale $\Lambda_\text{CFT}$, the renormalizable Yukawa couplings are
\be\label{eq:Ymarg}
W \supset Y^u_{ij} \,Q_i\, H_u\, \overline u_j+Y^d_{ij} \,Q_i\, H_d \,\overline d_j+Y^u_{33} \,Q_3 \,H_u \,\overline u_3+Y^d_{33} \,Q_3\, H_d\, \overline d_3
\ee
where $i,j=1,2$ and all the coefficients are taken to be order one. Renormalizable mixing terms between the third generation and the first two are forbidden by gauge invariance.  They will be generated by irrelevant operators as we explain below.

The third generation Yukawas are relevant in the CFT regime.  Below $\Lambda_\mr{CFT}$ they flow to order one fixed values. In contrast, the first two generation Yukawas become irrelevant because the Higgs fields acquire positive anomalous dimension. For energies, $v<E<\Lambda_\text{CFT}$ we find\footnote{The anomalous dimension is defined as $\Delta=1+\gamma/2$. For chiral primary operators it is related to the superconformal $R$-charge by $\gamma=3R-2$.}
\be\label{eq:runningy3}
Y_{ij}^u(E) =\left(\frac{E}{\Lambda_\mr{CFT}} \right)^\frac{\gamma_{Q_i}+\gamma_{u_j}+\gamma_{H_u}}{2} Y_{ij}^u(\Lambda_\mr{CFT})\,,
\ee
and a similar expression for $Y^d$. Defining the ratio between the exit scale and dynamical scale
\be\label{eq:epdef}
\epsilon \equiv \frac{v}{\Lambda_\text{CFT}}\,,
\ee
the suppression in the first two generation Yukawas at the exit scale becomes
\be\label{eq:runningy3}
Y_{ij}(v) = \epsilon^{\frac{\gamma_H}{2}} Y_{ij}(\Lambda_\mr{CFT}) \ll Y_{33}(v)\,.
\ee
We have neglected the perturbative anomalous dimensions for the first two generations. This dynamically generates a hierarchy between the first/second and the third generation Yukawa couplings. 

Next we consider the off-diagonal Yukawa interactions between the third and first/second generations. The lowest dimension operators allowed by gauge invariance are of the form
\be\label{eq:offY}
W\supset \frac{1}{\Lambda_*} \overline{\Sigma} \,Q_3\, H_u\, \overline u_{1,2}+\frac{1}{\Lambda_*} Q_{1,2}\, H_u\, \Sigma\, \overline u_3 + \ldots\,,
\ee
where $\Lambda_*$ is the scale at which these operators are generated. These lead to off-diagonal Yukawas after setting $\langle \Sigma \overline{\Sigma} \rangle=v^2$ at the exit scale. The RG evolution between $\Lambda_\text{CFT}$ and $v$ yields
\be\label{eq:runningy4}
Y^u_{i3}(v) = \frac{v}{\Lambda_*} \epsilon^{\frac{\gamma_{H_u}+\gamma_{Q_3}+ \gamma_{\overline \Sigma}}{2}}\;,\;Y^u_{3i}(v) = \frac{v}{\Lambda_*} \epsilon^{\frac{\gamma_{H_u}+\gamma_{u_3}+ \gamma_{\Sigma}}{2}}\,.
\ee

Note that the theory near the UV free fixed point contains two types of classically irrelevant operators: the $U(1)$ symmetry breaking term \eref{eq:ExplicitWU1} and the interactions \eref{eq:offY}. However, their IR fate is very different. As we showed before, the interaction $W_{\leavevmode\cancel{U(1)}}$ becomes relevant in the IR, driving the theory to a strongly coupled fixed point (where it becomes order one), while \eref{eq:offY} is irrelevant along the whole flow toward the fixed point. Such irrelevant perturbations do not modify the RG flow or the suppression of soft parameters. They become marginal only after the exit of the conformal regime.

Combining these results, we obtain the following the Yukawa textures at $v$:
\be\label{eq:Yu}
Y^u \sim \left( \begin{array}{ccc}
\epsilon^{\frac{\gamma_{H_u}}{2}} & \epsilon^{\frac{\gamma_{H_u}}{2}} & \xi_Q \,\epsilon^{\frac{\gamma_{H_u}}{2}}\\
\epsilon^{\frac{\gamma_{H_u}}{2}} & \epsilon^{\frac{\gamma_{H_u}}{2}} &  \xi_Q\,\epsilon^{\frac{\gamma_{H_u}}{2}}\\
\xi_u \,\epsilon^{\frac{\gamma_{H_u}}{2}}& \xi_u \,\epsilon^{\frac{\gamma_{H_u}}{2}}&  1
\end{array} \right),
\ee
with $\epsilon \ll 1$ defined in \eref{eq:epdef} and 
\be\label{eq:xidef}
\xi_Q \equiv \frac{v}{\Lambda_*}\,\epsilon^{\frac{\gamma_{\overline{\Sigma}}+\gamma_{Q_3}}{2}}\;,\;
\xi_u \equiv\frac{v}{\Lambda_*}\,\epsilon^{\frac{\gamma_{\Sigma}+\gamma_{\bar u_3}}{2}}\,.
\ee
A similar expression holds for $Y^d$. Choosing $\Lambda_*$ below the dynamical scale of the CFT and requiring negative $\gamma_{\overline{\Sigma}}+\gamma_{Q_3}$ and $\gamma_{\overline{\Sigma}}+\gamma_{\bar u_3}$ (as is the case in our model) gives $\xi_{Q, u} \gtrsim 1$.

For the model with the superpotential given in \eref{eq:ExplicitWU1}, the anomalous dimensions which determine the Yukawa couplings are $\gamma_{H_u}= \gamma_{H_d}=1$ and $\gamma_{Q_3}+ \gamma_{\overline \Sigma}=\gamma_{\bar u_3}+ \gamma_{\Sigma}=-3/2$. The flavor hierarchies between the third and second generations can be generated when
\be\label{eq:flavorh}
\frac{v}{\Lambda_\text{CFT}} \sim 10^{-4}\;,\;\frac{\Lambda_*}{\Lambda_\text{CFT}} \sim 10^{-1} - 10^{-2}\,.
\ee
This model does not explain why the first generation Yukawa is smaller than the second generation one.
However, this additional small Yukawa could arise by an accidental degeneracy of \eref{eq:Yu}, or by approximate flavor symmetries as in~\cite{Barbieri:1996ww}.  We have checked that by scanning over order one coefficients, we can reproduce the quark spectrum and the CKM matrix to a good approximation.

This ends the general analysis of our mechanism. The rest of the work is devoted to a study of its phenomenological consequences.

\section{Low Energy Phenomenology}\label{sec:SpectrumAndPheno}

Having explained the main features of our mechanism, we will now analyze the properties of the spectrum and Higgs sector and the parameter ranges which lead to a realistic low energy phenomenology.

\subsection{General Properties of the Spectrum}\label{subsec:genspectrum}

In this section, we discuss the features of the low energy spectrum in models which use the dynamics described in \S \ref{sec:TheModel}.  Supersymmetry breaking is communicated to the MSSM at the messenger scale $M$, where the operators
\be
c^2_{\t f}\int d^4 \theta \frac{X^\dag X}{M^2} \Phi^\dag_\mr{SM} \Phi_\mr{SM}\;,\;c_{\t g} \int d^2 \theta\,\frac{X}{M} \mathcal{W}_\alpha \mathcal{W}^\alpha\;,\;\ldots
\ee
are generated, where $X$ is a supersymmetry breaking spurion with $\vev{X} \supset \theta^2 F$, $\Phi_\mr{SM}$ is an MSSM matter superfield, $\mathcal{W}_\alpha$ is the field strength for an MSSM gauge group, and the factors of $c$ are model dependent coefficients.  These terms give sfermion and gaugino masses which are determined by the $F$-term of $X$. Supersymmetry breaking is external to the dynamics described in \S \ref{sec:TheModel}, and we do not constrain the soft UV boundary values, up to the assumptions on approximate symmetries required to suppress the differences given in \eref{eq:unsuppressed}. 

Typically in concrete models of supersymmetry breaking, the sfermion masses at the messenger scale are comparable for the three generations. On the other hand, sfermion and gaugino masses need not arise at the same order in $F/M$. This happens in many known cases.  For instance, an approximate $R$-symmetry or gaugino screening (which occurs for a wide class of gauge mediated models \cite{ArkaniHamed:1998kj}) can lead to subleading gaugino masses. We will assume that gauginos are around the TeV scale.  In principle the sfermions can be much heavier at the messenger scale, but we do require that 
\be
\widetilde{m}_{f} \sim c_{\t f}\frac{F}{M} \ll \Lambda_\text{CFT}
\ee
so that the conformal dynamics will be relatively unperturbed.

Generic sfermion masses will lead to flavor changing neutral currents (FCNCs). In our setup, flavor problems can be somewhat alleviated by having heavy enough sfermions, while also imposing some degree of degeneracy between the first two generations.\footnote{This is satisfied automatically if the mediation mechanism is flavor-blind.} In this case, $\widetilde{m}_{f_{1,2}} \gtrsim \mc {\mathcal O}(10 \mbox{ TeV})$ avoids dangerous FCNCs. On the other hand, there is a limit on how heavy the first two generations can be so that the third generation sfermion masses do not become tachyonic~\cite{ArkaniHamed:1997ab}. To account for this constraint, we include the dominant 2-loop contributions from the heavy states in the analysis of \S\ref{subsec:PhenoExampleSpect}. It would also be interesting to study models where the CFT dynamics alleviates such tachyonic contributions, allowing a more complete decoupling of the first two generation sfermions.

Once we enter the conformal regime, the soft masses for the third generation sfermions and Higgs fields are renormalized by the strong dynamics as described in \S \ref{sec:TheModel}, while the first two generation sfermions and gauginos are not appreciably modified. When evaluating the running of the soft parameters in the conformal regime we must consider that gauginos and first/second generation fields are continuously feeding supersymmetry breaking contributions into the third generation and Higgs fields~\cite{Nelson:2001mq}. For most of the viable parameter space, the dominant contribution comes from the gauginos, leading to finite contributions
\be\label{eq:mCFT}
\widetilde{m}^2_\text{CFT} \sim \frac{g_X^2}{16\, \pi^2} |M_3|^2
\ee
where $M_3$ is the MSSM gluino mass and $g_X$ is the gauge coupling of the weakly interacting $SU(3)_X$.

After escaping the conformal regime, we find a soft spectrum with $\widetilde{m}_{1,2}\sim \mathcal{O}\mbox{(few TeV)}$, $M_3 \sim {\mathcal O}(1 \mbox{ TeV})$, and small masses for the third generation squarks and Higgs fields. The masses for the light fields are then predominantly regenerated by gaugino mediation \cite{Kaplan:1999ac} between $v$ and the electroweak scale. The gauginos drive the stop mass to positive values.  For $v\gtrsim 50 \mbox{ TeV}$, this makes the up-type Higgs tachyonic and triggers electroweak symmetry breaking. The RG evolution will be studied explicitly below.

\subsection{Comments on the Higgs Sector}\label{subsec:higgs}

Next we discuss the interplay between the MSSM Higgs sector and our model.  First we consider the supersymmetric Higgs mass $\mu$ and the bi-linear supersymmetry breaking Higgs mass $b_\mu$. Our model contains a solution to the $\mu$ problem via the irrelevant interaction
\be\label{eq:muW}
W \supset \frac{1}{\Lambda_{\mu}} \Sigma\, \overline{\Sigma}\, H_u\, H_d\,.
\ee
(The tree level $\mu$ term $W \supset H_u H_d$ can be forbidden by symmetries.) The operator in \eref{eq:muW} can be generated by the same mechanism which produces the off-diagonal Yukawas in \eref{eq:offY}.  This is another interesting connection between flavor textures and the Higgs sector. Assuming this occurs, $\Lambda_{\mu} \sim \Lambda_*$ and no new scale is needed.\footnote{One can also imagine a different discrete symmetry such that  $\Sigma^3 H_u H_d$ is the lowest dimension operator which could generate an effective $\mu$ term.}  Taking into account the CFT suppression, the $\mu$ term at the exit scale becomes
\be\label{eq:mu}
\mu =\left( \frac{v}{\Lambda_*}\, \epsilon^{\gamma_{H}+\gamma_\Sigma}\right)\,v\,.
\ee
In this approach, $b_\mu$ is zero at $v$ and is generated radiatively as we run down to the weak scale.  In the leading log approximation,
\be
b_\mu \simeq -\frac{1}{16 \,\pi^2}\,\mu\left(6\,g_W^2 M_2 + \frac{6}{5}g_Y^2 M_1\right)\log\left(\frac{v}{m_W}\right),
\ee
where $M_1$ is the bino mass and $M_2$ is the wino mass. This solution to $\mu$ and $b_\mu$ can lead to REWSB.

For our model, $\gamma_{H}+\gamma_\Sigma=0$. Requiring $\mu \sim 100$ GeV and using the approximate values in \eref{eq:flavorh},
\be\label{eq:vmu}
v \sim 100\;\text{TeV},\;\Lambda_* \sim 10^4 - 10^5\;\text{TeV},\;\mbox{and }\Lambda_\text{CFT} \sim 10^6\;\text{TeV}.
\ee
While this is an attractive solution to the $\mu$ problem, when coupling our mechanism to a specific supersymmetry breaking model, there could be additional dynamics which explains $\mu/b_\mu$. In this case, it is not necessary to introduce \eref{eq:muW}, and the scales \eref{eq:vmu} could be different.

We now discuss the physical Higgs mass. Below the exit scale, the gluino mass will drive the stop mass positive, which in turn contributes negatively to $\widetilde{m}_{H_u}^2$. As long as the bino and wino masses are not too large, this will trigger electroweak symmetry breaking. Models with unified gauginos provide an example of successful REWSB.  The down-type Higgs soft mass will be generated though a combination of competing effects from the sbottom and the heavy first/second generations (which drive it negative), and the bino and wino (which drive it positive). 

Since the mechanism described in this work yields light stops and negligible $a$-terms, there is tension with a physical Higgs mass of order $125$ GeV, as currently hinted at by the LHC~\cite{ATLAS:2012ae}. Thus, a realistic model must include an additional source to raise the physical Higgs mass. In the simplest version of our construction, an NMSSM type extension does not solve this problem because the CFT makes the interaction $W \supset S\, H_u\, H_d$ (with $S$ the extra singlet in the NMSSM) irrelevant.  This leads to a negligible increase in the physical Higgs mass.  One option beyond singlet extensions would be to add ``non-decoupling $D$-terms''~\cite{Maloney:2004rc} below the exit scale. While we do not attempt to embed this or other mechanisms into our model, we see no fundamental obstruction.  The validity of our conclusions require that this additional module does not lead to appreciable shifts for any of the soft masses.

\subsection{An Example Spectrum}\label{subsec:PhenoExampleSpect}

In order to perform a concrete analysis, we will work in the context of a model with unified gaugino masses. We will also assume that the mediation of supersymmetry breaking respects custodial symmetry and a ``charge conjugation" symmetry between $Q$ and $\overline{u},\,\overline{d}$, \emph{i.e.}, $\widetilde{m}_{Q_3}^2 = \widetilde{m}_{\overline{u}_3}^2 = \widetilde{m}_{\overline{d}_3}^2$.  For example, both of these assumptions are well approximated by models of minimal gauge mediation. The following analysis demonstrates in a concrete setup the viability of the mechanism for splitting the third generation from the first and second.  The techniques presented here can be applied to a wide class of supersymmetry breaking scenarios.  

Given this framework, the spectrum is determined by choosing a gluino mass and solving the RG equations with the boundary condition at the scale $v$ which the third generation and Higgs soft masses are given by\eref{eq:mCFT}. While there is an incalculable order one coefficient, we find that such effects are small in the regime of interest. If we also assume the solution to the $\mu$ problem proposed in \S \ref{subsec:higgs}, the exit scale is fixed at $v \sim \mc O(100)\; \text{TeV}$. The model is then very predictive: all we need to specify are the messenger scale, gaugino and first/second generation masses.

As an example, we find the viable spectrum presented in Table \ref{tab:viableSpectrum}, with first/second generation sfermion masses chosen to be $5$ TeV. We have assumed an additional contribution to the Higgs quartic from a coupling $g_\mr{new}$ so that
\be\label{eq:Xi}
m_Z^2 = \frac{g_Z^2}{2} \left(\vev{H_u}^2 +\vev{H_d}^2\right) \ \quad \longrightarrow\quad  \Xi^2 \equiv  \frac{g_Z^2+g_\mr{new}^2}{2}\left(\vev{H_u}^2 +\vev{H_d}^2\right).
\ee
in all tree-level MSSM expressions for electroweak symmetry breaking and the Higgs sector.  In our numerical analysis below, we will take $\Xi \simeq 150 \mbox{ GeV}$.  As discussed in \S \ref{subsec:higgs}, this could in principle arise from a non-decoupling $D$-term --- we are agnostic about its source and find that this leads to a small change for all the parameters except for the physical value of the CP even Higgs masses.  This yields a Higgs mass of $105$ GeV at tree-level which (given the stop masses in Table \ref{tab:viableSpectrum}) will lead to a mass consistent with $125$ GeV once loop corrections are taken into account.  This point is also consistent with the relevant experimental bounds considered in \S \ref{subsec:ExploreParamSpace} below.   This demonstrates the viability of our mechanism.

\begin{table}[h!!!]
\begin{center}
\begin{tabular}{|c|c|c|c|}
\hline
 &&& \\[-12pt]
$v$ & $M_3$ & $M_2$ & $M_1$ \\
 &&& \\[-12pt]
 \hline
  &&& \\[-12pt]
350 TeV & 2.5 TeV & 1.0 TeV & 530 GeV \\
 &&& \\[-12pt]
 \hline
  \hline
 &&& \\[-12pt]
 $\widetilde{m}_{3}^2$ &$\widetilde{m}_{1,2}^2$  & $\widetilde{m}_{H_u}^2$ & $\widetilde{m}_{H_d}^2$ \\
  &&& \\[-12pt]
\hline
 &&& \\[-12pt]
  (1.2 TeV)$^2$ & (5 TeV)$^2$ & $-(220\mbox{ GeV})^2$ & (300 GeV)$^2$\\
   &&& \\[-12pt]
  \hline
   \hline
    &&& \\[-12pt]
$\mu$ & $b_{\mu}$ & $M_A$ & $\tan \beta$\\
 &&& \\[-12pt]
\hline
 &&& \\[-12pt]
220 GeV & - 0.030 GeV$^2$ & 135 GeV & 4.2\\
\hline
\end{tabular}
\end{center}
\caption{\small An example set of consistent parameters with the solution to the $\mu$ problem given in \eref{eq:mu}.  We have assumed gaugino mass unification and to good approximation $\widetilde{m}_{Q_i}^2=\widetilde{m}_{u_i}^2=\widetilde{m}_{d_i}^2=\widetilde{m}_i^2$ at low energies.  We find that the tree-level value of the Higgs mass is $ \simeq 105\mbox{ GeV}$ which is consistent with 125 GeV when loop corrections are taken into account.}
\label{tab:viableSpectrum}
\end{table}

\subsection{Exploring the Parameter Space}\label{subsec:ExploreParamSpace}

\begin{figure}[ttt]
\begin{center}
\includegraphics[width=0.47\textwidth]{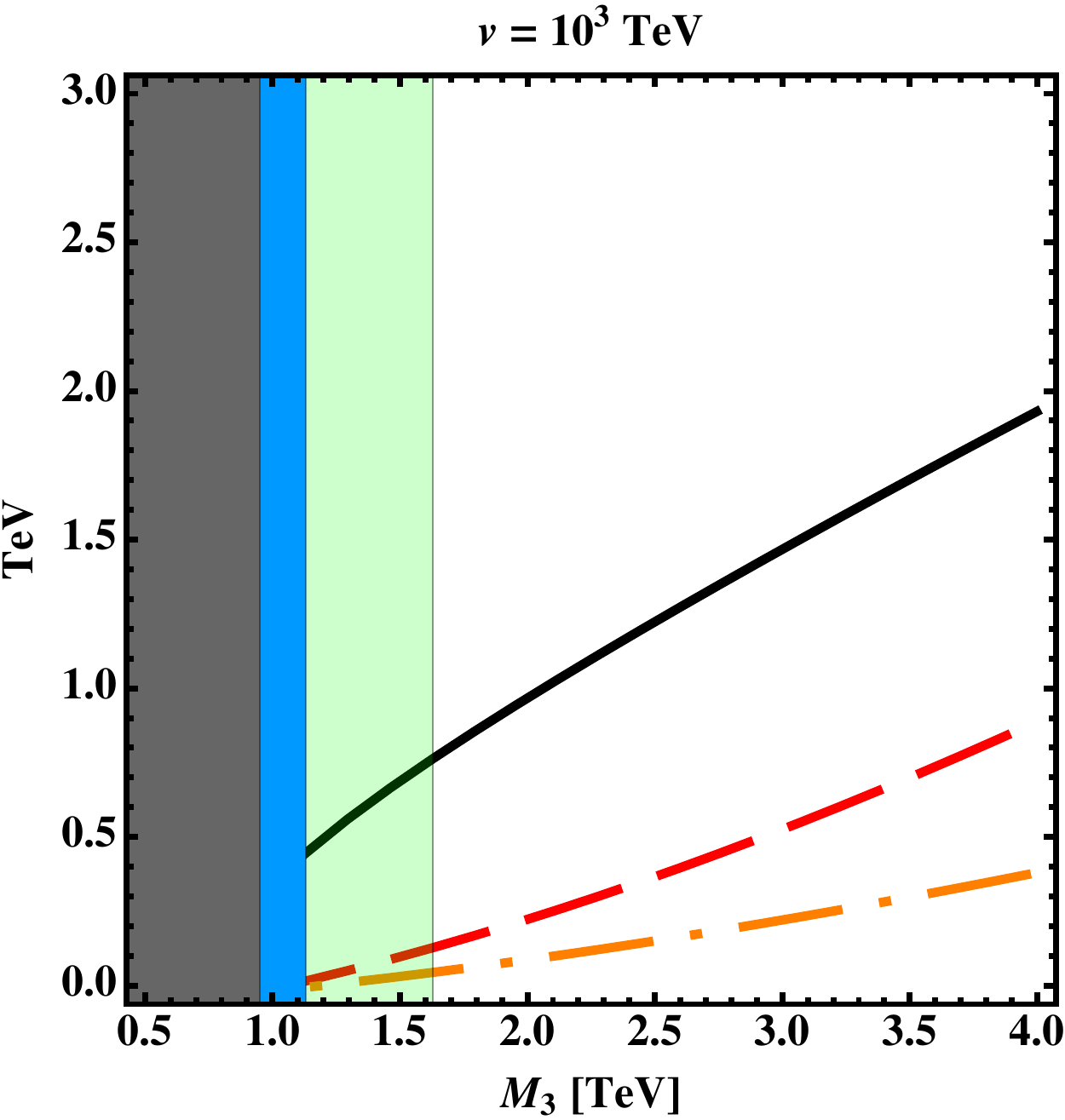}\quad\quad
\includegraphics[width=0.47\textwidth]{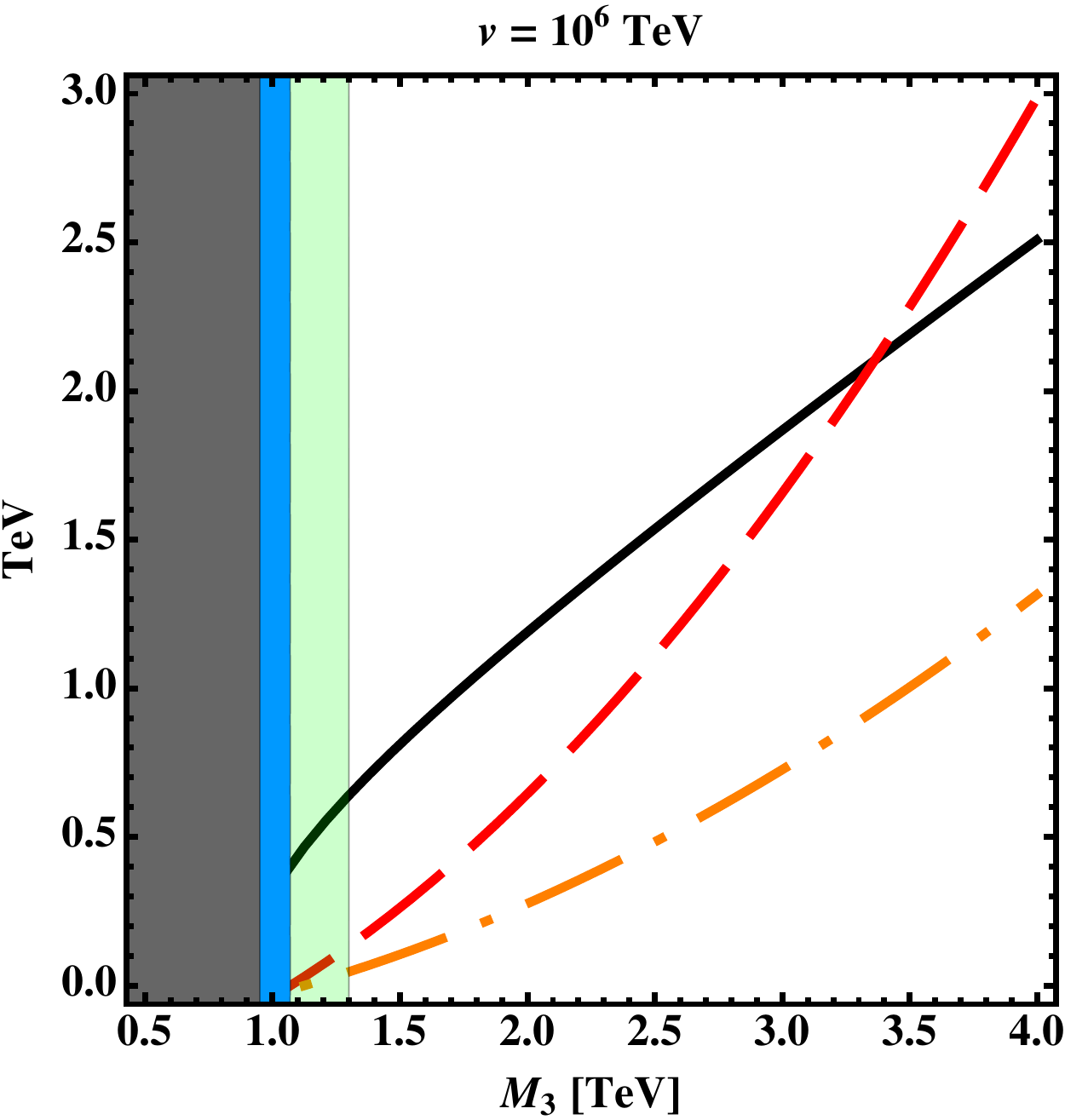}
\end{center}
\caption{\small Low energy spectrum for a model with unified gaugino masses, for $v = 10^3\mbox{ TeV}$ (\emph{left}) and $v = 10^6\mbox{ TeV}$ (\emph{right}). The curves represent $\widetilde{m}_{Q_3}\simeq \widetilde{m}_{u_3} \simeq \widetilde{m}_{d_3}$ [black, solid], $m_A\, (\mbox{with }\tan\beta = 2)$ [red, dashed], $m_A\, (\mbox{with }\tan\beta = 10)$ [orange, dotted-dashed].  (Only $m_A$ has a strong dependence on $\tan\beta$.)  The first/second generation squark masses are at 5 TeV.  The opaque grey region is excluded due to tachyonic third generation squarks.  The opaque blue region is excluded by requiring radiative electroweak symmetry breaking.  The light translucent green region is excluded due to the LEP bound on the $A$ mass.  Both of these regions are plotted for the $\tan \beta = 2$ case. We fix $\Xi = 150 \mbox{ GeV}$ and for simplicity do not attempt the model dependent task of reproducing the Higgs mass for all points in this plot.}
\label{fig:spectra}
\end{figure}

In this subsection we will briefly explore the possible range of predictions for the soft mass spectrum.  In order to do this we will relax the relationship between the $\mu$ term and $v$ given in \eref{eq:mu}.  Noting that in our concrete model the coupling $W \supset H_u H_d$ is exactly marginal at the fixed point, one can in principle generate $\mu$ and $b_{\mu}$ using an unrelated mechanism at scales above $\Lambda_\mr{CFT}$.  We can thus take $v$ and $\tan \beta$ as free parameters and explore the resultant phenomenology.  In Figure~\ref{fig:spectra} we have plotted the low energy values of $\widetilde{m}_{Q_3}^2\simeq \widetilde{m}_{u_3}^2 \simeq \widetilde{m}_{d_3}^2$ [black, solid] , $m_A\, (\mbox{with }\tan\beta = 2)$ [red, dashed], and $m_A\, (\mbox{with }\tan\beta = 10)$ [orange, dotted-dashed] for two choices of $v$ as a function of the gluino mass.  The mass of the $A$ is the only parameter with a strong dependence on $\tan \beta$.  As in \S \ref{subsec:PhenoExampleSpect}, we assume that supersymmetry breaking respects $\widetilde{m}_{Q_3}^2 = \widetilde{m}_{\overline{u}_3}^2 = \widetilde{m}_{\overline{d}_3}^2$ to a good approximation.

In order to generate this plot, we use the RG equations for the MSSM to flow from $v$ to the weak scale including the leading 2-loop contributions from the first and second generation sparticles which we fix at 5 TeV.  It is this choice which causes the third generation squarks to become tachyonic for small gluino masses.  This is the excluded region plotted in opaque grey in Figure~\ref{fig:spectra}.  The opaque blue region is excluded due to a lack of REWSB (these conditions are unchanged from the MSSM).   The light translucent green region is excluded due to the LEP bound on the $A$ mass\footnote{This exclusion is highly dependent on $\tan\beta$.  Furthermore, one could imagine a model where $H_d$ is not a part of the CFT.  It would have a large mass and the model would generically be in the decoupling limit.} \cite{Schael:2006cr}.  This constraint is cut-off by the kinematic reach of LEP for the process $e^+ e^- \rightarrow h\, A$.  For $m_h = 125\mbox{ GeV}$ $(115\mbox{ GeV})$, this implies that $m_A \gtrsim 90\mbox{ GeV}$ ($m_A \gtrsim 100\mbox{ GeV}$).  As a conservative estimate, we impose $m_A > 100 \mbox{ GeV}$ in Fig.~\ref{fig:spectra}.  We have not included the bino, wino, and first/second generation soft masses  in Fig.~\ref{fig:spectra} since they are unaffected by our mechanism up to small effects due to off-diagonal Yukawa couplings and 2-loop diagrams. 

The bounds for the $\tan \beta = 10$ case are $M_3 > 0.85\, (0.92) \mbox{ TeV}$ to avoid having tachyonic squarks, $M_3 > 1.2\, (1.1) \mbox{ TeV}$ for REWSB and $M_3 > 2.1 \,(1.9) \mbox{ TeV}$ for the $A$ mass, given $v = 10^3 \,\left(10^6\right) \mbox{ TeV}$.  Note that the LHC also places strong bounds on $m_A$ from searches for di-tau resonances \cite{Chatrchyan:2012vp}.  In fact, the LHC excludes the range $120 \mbox{ GeV}\lesssim m_A \lesssim 220 \mbox{ GeV}$ for $\tan\beta = 10$ in the context of the MSSM (with $\Xi = m_Z$).   We do not show these constraints in Fig.~\ref{fig:spectra} since the excluded regions are for the $\tan \beta = 2$ example.

Recall that achieving a Higgs mass of 125 GeV requires physics beyond the simple model proposed here.  Hence, we will only make a few comments about the mass of the Higgs. First, we note that $A$ is light in the region of parameter space with the lightest squarks and gluino, which can have a non-trivial impact on the mass and couplings of the $h$.  In the pure MSSM, this manifests as a dependence on both $\tan \beta$ and the Higgs mixing angle $\alpha$ for the Higgs couplings (for a review, see \cite{Djouadi:2005gj}). More generally, the dependence of the Higgs couplings on $m_A$ is model dependent. It would be interesting to develop a realistic model for the Higgs sector based on our general mechanism, where this and related questions could be addressed in detail.

In generating Fig.~\ref{fig:spectra}, we took $\Xi = 150\mbox{ GeV}$, see \eref{eq:Xi}; we find only mild sensitivity to the choice of $\Xi$.  When $m_A < \Xi$, $m_h \simeq m_A \cos(2\,\beta)$, independent of $\Xi$.  For the choice $\tan \beta = 10$, the one-loop corrections from the stops are approximately of the right size to generate a Higgs mass of 125 GeV in the allowed window $100 \mbox{ GeV} \lesssim m_A \lesssim 120 \mbox{ GeV}$.  For larger values of $m_A$, the tree-level contribution to the Higgs boson mass would be set by $\Xi$, which could be carefully chosen to reproduce the desired result.  Alternatively, one could attempt to alter the Higgs quartic with a different mechanism than the one captured by our parameter $\Xi$.

Since we have a splitting between the first/second and third generation squarks, we must worry about FCNC effects induced by rotating the Yukawa matrices of \eref{eq:Yu} to the physical basis.  To leading order in $\epsilon$, the relevant 1-3 and 2-3 mixing is given by $\delta_{i3} \sim \epsilon^{\gamma_{H}/2}\xi$, where $\delta_{ij} \equiv \widetilde{m}^2_{ij}/\mr{max}(\widetilde{m}^2_{ii},\,\widetilde{m}^2_{jj})$. Assuming some degree of degeneracy between the first and second generations, negligible $a$-terms, and an absence of CP violating phases (as in minimal gauge mediation), the strongest flavor bound is from $\left(\delta^d_{13}\right)_{\mr{LL}=\mr{RR}} \lesssim 5\times 10^{-3}$ \cite{Buchalla:2008jp}.  There are also potential constraints from $b\rightarrow s\,\gamma$ and $B_s\rightarrow \mu^+\mu^-$ which are sensitive to model dependent choices, such as details of the chargino sector.  Overall, we find no impediment to accommodating these constraints in our model.  

\begin{figure}[ttt!!]
\begin{center}
\includegraphics[width=0.5\textwidth]{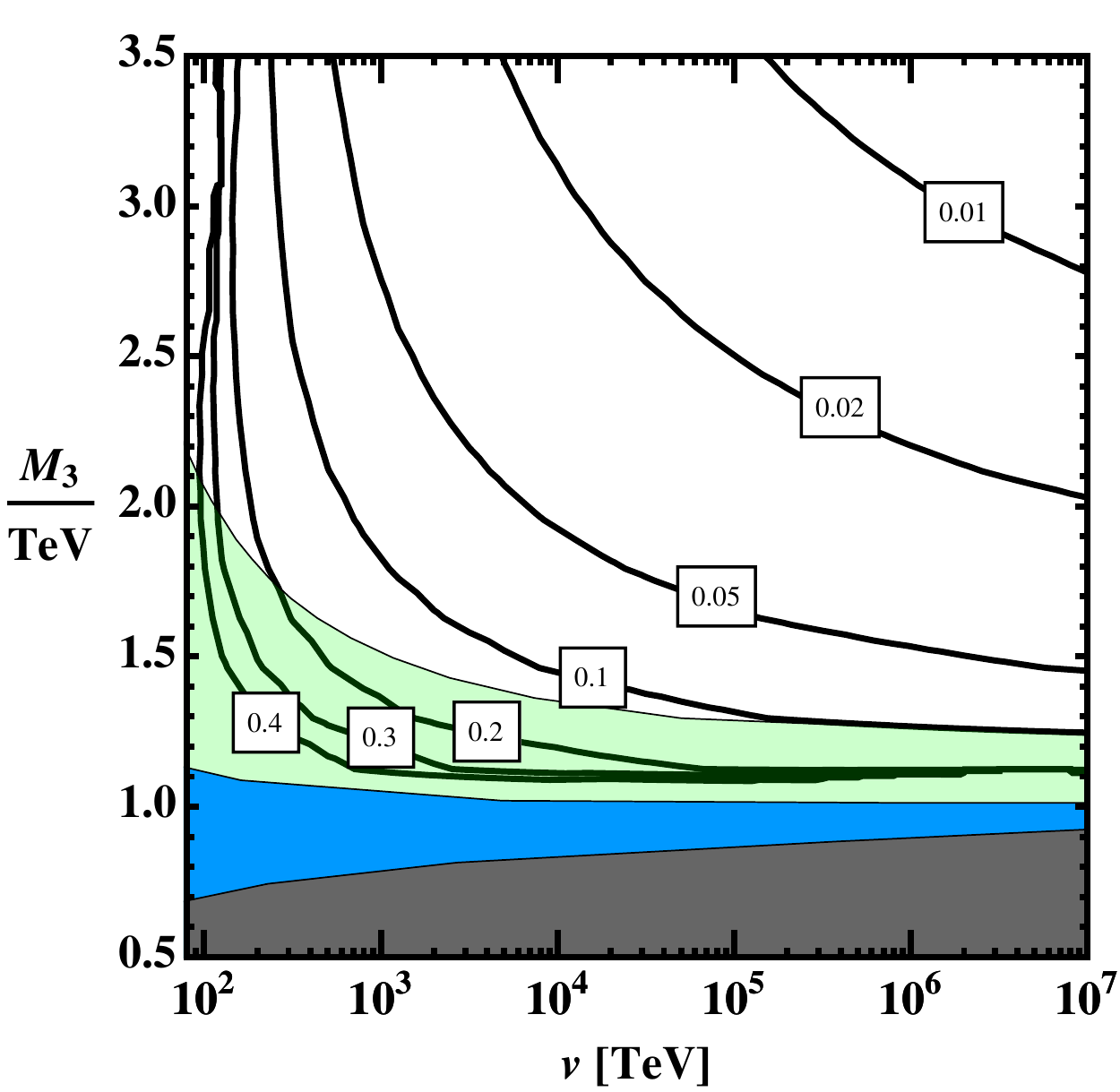}
\end{center}
\caption{\small We plot contours of the ``fine tuning" parameter $\Delta^{-1}$ in the $M_3$ versus $v$ plane.  We make the same assumptions as in Fig.~\ref{fig:spectra} with first/second generation squark masses at 5 TeV and $\tan \beta = 2$.  The solid grey region is excluded due to tachyonic third generation squarks, the solid   blue region is excluded due to a lack of REWSB and the green translucent region is excluded due to the LEP bound on the $A$ mass.}
\label{fig:FineTuning}
\end{figure}

Finally, let us briefly discuss the contributions to fine tuning which result from our mechanism.  The problem of naturalness is related to the question of curvature in the symmetry breaking direction --- it is a one-dimensional problem for a Higgs field $H$ as in the standard model with a potential
\be
V = m_H^2 |H|^2 + \lambda |H|^4.
\ee
When $\vev{H} \neq 0$, the physical Higgs mass squared $m_h^2 = - 2\,m_H^2$.  One simple measure of fine tuning, advocated in \cite{Kitano:2005wc,Papucci:2011wy}, is then
\be\label{eq:FineTuningMeasure}
\Delta^{-1} \equiv -2 \frac{\delta m_H^2}{m_h^2} = -2 \frac{\widetilde{m}_{H_u}^2}{m_h^2},
\ee 
where in the last equality we are interested in the contribution to the Higgs soft mass in our model.

In Fig.~\ref{fig:FineTuning} we have plotted contours of $\Delta^{-1}$ from our dynamics in the $M_3$ versus $v$ plane.  The most important assumption from the point of view of fine tuning is the gaugino mass spectrum.  We have also plotted the region which is excluded due to tachyonic third generation squarks in solid grey, a lack of REWSB in solid blue, and the LEP bound on the $A$ mass for $\tan \beta = 2$ in translucent green.  We see that there is an allowed region with $\Delta^{-1}\simeq \mathcal{O}(10\%)$ where $v\simeq 10^2 \mbox{ TeV}$ and $M_3 \gtrsim 2.5 \mbox{ TeV}$.  We note that in a complete model which addresses the physical Higgs boson mass, there may be additional sources of fine tuning.

\section{Conclusions and Future Directions}\label{sec:concl}

In this work we have presented a mechanism which acts as an attractor for the more minimal supersymmetric standard model and radiative electroweak symmetry breaking, while also generating the hierarchical structure of the quark Yukawa matrix. We have presented the simplest realization, which is accomplished by adding a new $SU(3)$ gauge group under which the third generation quarks are charged. The model flows to a strongly interacting fixed point where these quarks acquire order one negative anomalous dimensions, while the Higgs gets a positive anomalous dimension. The mechanism applies to generic supersymmetry breaking scenarios, as long as appropriate symmetries ensure that the combinations of masses in \eref{eq:unsuppressed} are small. It also leads to a simple solution of the $\mu$ problem. For concreteness we analyzed the low energy phenomenology starting from unified gaugino masses, finding a natural supersymmetry spectrum with split families.

It would be interesting to build a fully realistic model based on this mechanism. The main points which need to be addressed are unification (which has been explicitly broken here by the extra matter charged under $SU(3)$) and the generation of a realistic physical Higgs mass. This motivates extending our approach by having two copies of the full SM gauge group, instead of just the $SU(3)$ group. One of the nodes will then become strongly coupled, leading to the properties analyzed here. In this context, the NMSSM can naturally become part of the strong dynamics, and unification is in principle possible. 

The mechanism itself can also be improved in different directions. Here we had to assume that certain approximate symmetries of the supersymmetry breaking sector were forbidding the combinations of soft masses given in \eref{eq:unsuppressed}. In particular, combinations proportional to $U(1)_Y$ can not be screened. This can be avoided if $U(1)_Y$ is embedded into a larger gauge group for the duration of the conformal regime. One possibility would be to weakly gauge the custodial $SU(2)$. In this case, the only combinations which are not sequestered are $\widetilde m_\Sigma^2-\widetilde m_{\overline\Sigma}^2$ and $\widetilde m_{Q_3}^2-\widetilde m_{\overline{Q}_3}^2$, where $\overline{Q} = (\overline{u},\,\overline{d})$ --- both of these combinations can be suppressed by imposing a discrete symmetry. This can lead to a stronger attractor mechanism.  

For models which realize this stronger attractor, there is a novel possibility of decoupling the first/second generation squarks beyond the bound of \cite{ArkaniHamed:1997ab}.  If $\widetilde{m}_{1,2} \gg v$, at scales below $\widetilde{m}_{1,2}$ there will be a quadratically divergent contribution to the stop masses at 2-loops and the Higgs mass at 1-loop (which is proportional to small Yukawa couplings).  If it is possible to construct a CFT which would be strong enough to suppress these quadratic divergences, the contribution to the mass from these effects will be schematically given by $y^2/(16\, \pi^2)\, v^2$ for the Higgs soft mass squared and $g_C^4/(16\,\pi^2)^2\, v^2$ for the stop soft mass squared.  For $v\simeq 50\mbox{ TeV}$, these contributions are small enough to not destabilize our mechanism.  Hence, the flavor problem could be completely decoupled in these models.

It may also be possible to find a microscopic realization where the CFT and the sector which breaks supersymmetry are part of the same dynamics. In this setup, the exit scale $v$ would be related to the scale of supersymmetry breaking. This may be done at the level of the superpotential, or by destabilizing some of the flat directions of the CFT.  Exploring a concrete supersymmetry breaking sector which minimizes the mass differences in \eref{eq:unsuppressed} would also be an interesting avenue for future work. 

If nature cares about naturalness, it is plausible that the dynamics between the weak scale and Planck scale could be highly non-trivial.  We have demonstrated that coupling the supersymmetric standard model to a new strongly coupled conformal sector can give rise to the flavor hierarchies and the more minimal spectrum.

\section*{Acknowledgements}

We thank J.~Wacker for collaboration at early stages of this work.  We also thank I.~Bah, S.~El Hedri, T.~Gherghetta, D.~Green, H.~Haber, S.~Kachru,  J.~March-Russell, and M.~Peskin for useful conversations and N.~Craig for detailed comments on the manuscript.
This work was supported by the US Department of Energy under contract number DE-AC02-76SF00515.  
%


\newpage

\bibliographystyle{JHEP}
\renewcommand{\refname}{Bibliography}
\addcontentsline{toc}{section}{Bibliography}
\providecommand{\href}[2]{#2}\begingroup\raggedright

\end{document}